\documentclass[aps,prl,reprint,groupedaddress]{revtex4-1}

\usepackage{hyperref}
\usepackage{amsthm}
\usepackage{amsmath}
\usepackage{amsfonts}
\usepackage{latexsym}
\usepackage{color}
\usepackage{amssymb}
\usepackage{mathrsfs}
\usepackage{graphicx}
\usepackage{bm}
\usepackage{txfonts}
\usepackage{url}
\usepackage{ulem}
\begin{document}

\title{Spontaneous stable rotation of flocking flexible active matter}
\author{Gaoxiao Jiang}
\author{Zhihong You}
\author{Rui Ma}
\email{ruima@xmu.edu.cn}
\author{Chenxu Wu}
\email{cxwu@xmu.edu.cn}
\affiliation{Fujian Provincial Key Lab for Soft Functional Materials Research, Research Institute for Biomimetics and Soft Matter, Department of Physics, College of Physical Science and Technology, Xiamen University, Xiamen 361005, People's Republic of China}
\date{\today}
\begin{abstract}
 In nature, active matter, such as worms or dogs, tend to spontaneously form a stable rotational cluster when they flock to the same food source on an unregulated and unconfined surface. {In this paper we present an $n$-node flexible active matter model to study the collective motion due to the flocking of individual agents on a two-dimensional surface, and confirm that there exists a spontaneous stable cluster rotation synchronizing with a chirality produced by the alignment of their bodies under the impetus of the active force.} A prefactor of 1.86 is obtained for the linear relationship between normalized angular velocity and chirality. The angular velocity of such a rotation is found to be dependent on the individual flexibility, the number of nodes in each individual, and the magnitude of the active force. The conclusions well explain the spontaneous stable rotation of clusters that exists in many flexible active matter, like worms or {dogs}, when they flock to the same single source.
\end{abstract}

\maketitle

\section{Introduction}

Since the publication of the Vicsek model~\cite{Vicsek}, studying the collective motion of self-propelled agents has become an important topic in physics~\cite{Review_Liquid,Review_Complex,Review_Phenomenon,Review_Active}. Emergence of rotating structures from the collective motion of active agents is a fascinating phenomenon observed in nature, from swimming bacteria in the micro-meter scale~\cite{bacteria_vortex1,bacteria_vortex2,bacteria_vortex3,bacteria_vortex4}, to flocks of plant-animal worms~\cite{worms_vortex}, processionary caterpillars~\cite{processionary_caterpillars_vortex}, ants~\cite{ant_vortex}, tadpoles~\cite{tadpoles_vortex}, {plankton and fish in the meters scale~\cite{daphnia_vortex,fish_vortex1,fish_vortex2}. Such an emergent collective motion can be classified into swarm, polarized and milling (vortex) states\cite{61/10.1371/journal.pcbi.1002915} with the milling one typically associated with confined geometry~\cite{emergence_vortex1,emergence_vortex2}, orientational asymmetry by introducing a blind angle\cite{51PhysRevE.100.012609,41Cheng_2016,03PhysRevE.78.011913} or external field~\cite{emergencevortex3,Quincke_vortex3,Experiment_ElectricDriven2,Experiment_MagneticDriven,mag_roller__vortex1,mag_roller__vortex2,mag_roller__vortex3}. In the presence of an alternating electric field, experiments found that spherical particles with dipole moment can be self-propelled and form a transient/stable rotating cluster~\cite{emergencevortex3,Quincke_vortex3}. Similar phenomena were also observed with magnetic rollers or dielectric colloidal particles~\cite{Experiment_ElectricDriven2,Experiment_MagneticDriven,mag_roller__vortex1,mag_roller__vortex2,mag_roller__vortex3,Simulation_Star}, among which hydrodynamic interaction plays an important role.

By introducing one or a group of flexible structures into the matrix of active matter, the flexible structure can demonstrate spontaneous stable rotation. It was reported that when an aster-like structure with flexible arms is placed in the matrix of self-propelled particles, the arms can bend and form a spiral structure and spontaneously rotate under the collective collision of the active particles~\cite{Simulation_Star}. When placing long bacteria in a bath of short bacterial, the long bacteria were found to fold themselves under the collision of the short bacteria, and wrap around into a clockwise or anticlockwise coil, or double coils with both types~\cite{ExperimentLongBacteria}. In these studies, the rotation of a large flexible structure is typically associated with chiral symmetry breaking due to its interaction with a population of small hard particles.

If the active particles are intrinsically chiral, the large-scale rotation exhibits novel phenomena\cite{self_chiral}. For instance, a collection of self-spinning particles can form topologically protected edge flows that circulate around the boundary\cite{edge__flow1,edge__flow2}. Simulations have found that such a dense suspension of L-shaped particles exhibits a chirality-triggered oscillation phase, in which transient vortices are assembled and disassembled in a periodic manner~\cite{Shape_L}.

{Here it should be noted that if an attractive interaction, like Morse potential\cite{03PhysRevE.78.011913,21STROMBOM2011145,11Lukeman2009,04Hindes2021,02PhysRevLett.96.104302,01PhysRevE.63.017101}, exists in a dense suspension of active spherical particles, even in the absence of a confining boundary, a stable vortex (milling) structure can be obtained. The rotational stability due to perturbation was also discussed under the assumption of such  kind of milling\cite{71doi:10.1098/rsif.2014.1362}}.

In nature, even flexible individual agents that are achiral in their natural form, like worms~\cite{worms_vortex}, ants~\cite{ant_vortex}, can spontaneously form a stable rotating cluster once flocking together even under no spatial constraints in the absence of hydrodynamic interactions. {Such a phenomenon does not need an attractive interaction among individual agents, and the driving force is the information of a specific position where they would like to flock to due to something (such as food) there they are interested in.} Milling has been observed in colonies of bacteria due to the alignment pressure generated by dividing cells in a crowded environment ~\cite{Bacteria_Orientation,Bacteria_OrientationInVortices,Bacteria_Vortices,Bacteria_Vortex}. The mechanism of how a group of achiral flexible active particles, when they flock for a single source, form a stable rotating cluster spontaneously without any spatial confinement and hydrodynamic interactions remains to be elucidated.

In this paper, as a theoretical attempt, we present an $n$-node flexible active matter model to study the collective motion of a group of flocking achiral but flexible particles. It is found that a spontaneous stable rotation of the cluster can be generated as a result of the self-organization of the flexible individuals into a chiral structure.

\begin{figure}[h]
\includegraphics[width=250pt]{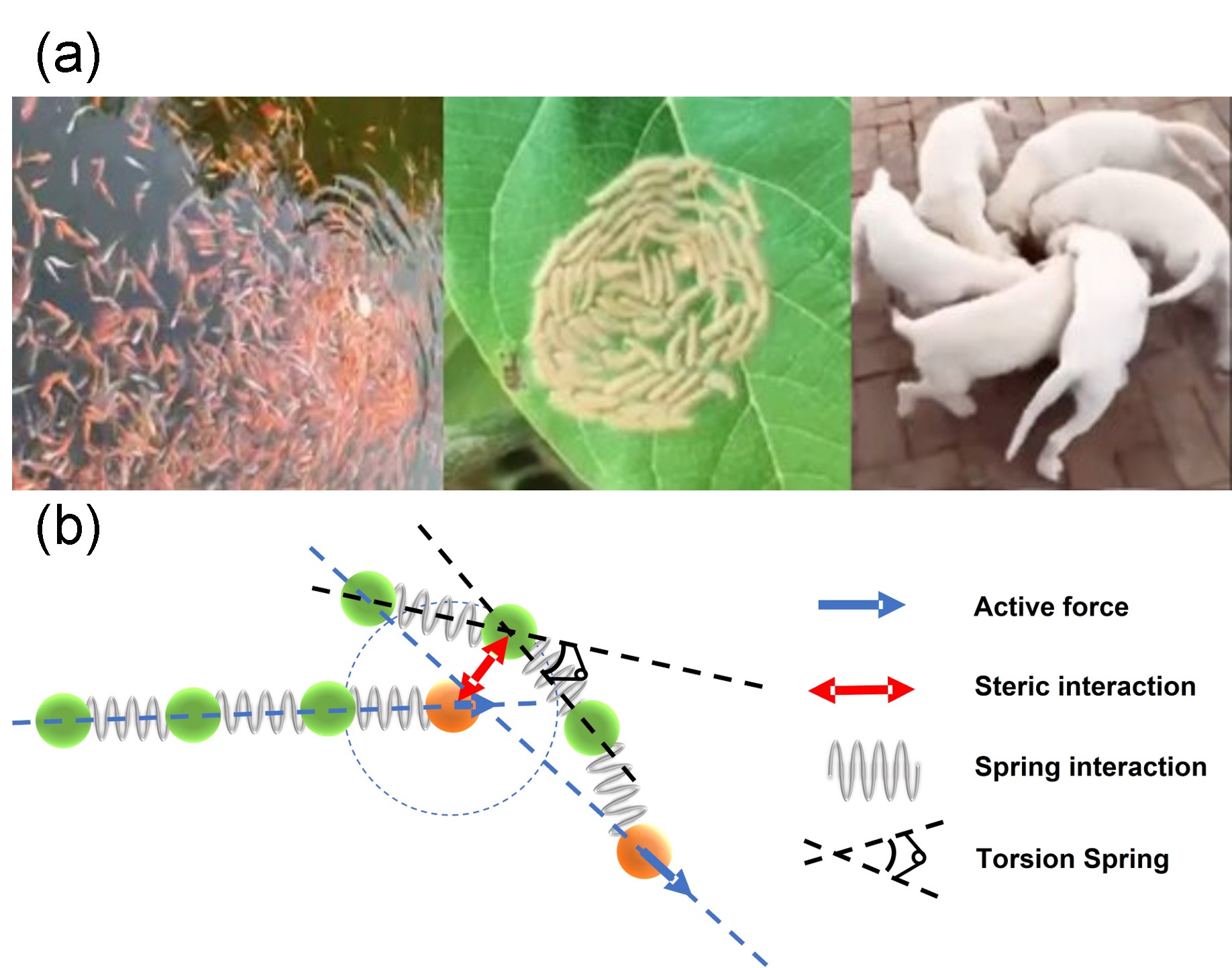}

	\caption{(a) Examples of spontaneous stable rotation of active matter due to flocking {~\cite{video1,video2,video3}} and (b) schematic of an individual consisting of $n$ nodes connected by springs with an active head.}
	\label{model}
\end{figure}

\section{Model of chain-like active matter}

Our model is designed to simulate the collective behaviors of flocking active matter on a two-dimensional surface, as shown in Fig.~\ref{model}(a). Each individual is modeled as an $n$-node (with node size { $ d=2^{1/6}\sigma $}) flexible chain bonded by harmonic springs, with its head labeled as the $1$-st node and its tail as the $n$-th node (Fig.~\ref{model}(b)).  A total number of $N$ individuals are placed in a square of length $L$ with periodic boundary conditions implemented at the four edges. The individuals tend to flock towards a specific point due to some reason such as a food source discovered there, and the system is assumed to be in an over-damped state during the flocking process with the dynamics of the $\alpha-$th node of the $i-$th individual governed by
\begin{equation}
	\label{equ:ld}
	\xi\partial{\mbox{\boldmath $r$} }_{i,\alpha}/\partial t=  \delta_{\alpha 1}f_i(N_i)\makebox{\boldmath $p$} _i -  \nabla_{r_{i,\alpha}}(E^{r}_{i,\alpha}+E^b_{i}+E^s_{i})+\xi\sqrt{2D}{\bm \theta}(t),
\end{equation}
where $\xi$ is the friction coefficient, ${\bm r}_{i,\alpha}$ is the position of the $\alpha$-th node in the $i$-th individual, $\xi {\sqrt{2D}\bm \theta}(t)$ is the Gaussian white noise, denoting the diffusion term, which satisfies $<{\bm \theta}(t)>=0$ and $<{\bm \theta}(t) {\bm \theta}(t')>=\delta(t-t')$. Hereafter, the Roman letters and the Greek letters are used to label the individual and the node in an individual, respectively.The first term on the right hand side of Eq.~(\ref{equ:ld}) is the active force with its magnitude $f_i$ inversely proportional to the number $N_i$ of nodes within a circle of a cutoff radius $R_{\rm c}$, but proportional to the number of nodes $n$ in each chain

\begin{equation}
	\label{equ:Ni}
	f_i = nf_{\rm max}/N_i.
\end{equation}
{This is based on the assumption that agents with bigger size are more powerful.} Here $f_{\rm max}$ denotes the maximum active force generated by the individual head, i.e., the first node for a single node individual.  The direction ${\bm p}_i$ of the active force is the unit vector pointing from tail to head, i.e.,
\begin{equation}
	\label{equ:p}
	\mbox{\boldmath $p_i$}=\frac{\mbox{\boldmath $r$}_{i,1}-\mbox{\boldmath $r$}_{i,n}}{||\mbox{\boldmath $r$}_{i,1}-\mbox{\boldmath $r$}_{i,n}||}.
\end{equation}
 Our model differs from previous ones in that the direction of the active force is chosen to point from tail to head rather than one just undergoing Brownian rotation in the traditional ABP models. Such a choice more accurately reflects the intelligence of individuals which would, besides the interaction with its surrounding during the flocking process, more actively steer and control their body based on the orientation of their own, rather than being simply and passively influenced by the collective velocity of its neighbors, as assumed in the classical Vicsek model~\cite{Vicsek}. The second term in Eq.~(\ref{equ:ld}) describes the interaction force between nodes. The steric interaction between node $\alpha$ in the $i$-th individual and node $\beta$ in the $j$-th individual is described by a repulsive energy $U_{WCA}(r_{i,\alpha}^{j,\beta})$ depending on the distance $r_{i,\alpha}^{j,\beta}=|{\bm r}_{i,\alpha}-{\bm r}_{j,\beta}|$ between them. Here $U_{WCA}(r)$ is the Weeks-Chandler-Anderson (WCA) potential energy given by
\begin{eqnarray}
	\label{equ:UWCA}
	U_{WCA}(r) =
\begin{cases}
	4\varepsilon[(\sigma/r)^{12}-(\sigma/r)^6]+\varepsilon       & r< 2^{1/6} \sigma  \\
	0    						        &  otherwise, \\
\end{cases}
\end{eqnarray}
where $r=2^{1/6}\sigma$ is a cutoff length, and $\varepsilon$ is the strength of the steric interaction. The total steric energy experienced by the $\alpha$-th node in the $i$-th individual reads
\begin{equation}
E^{r}_{i,\alpha} = \sum_{j=1}^N \sum_{\beta=1}^{n}U_{WCA}(r_{i,\alpha}^{j,\beta}).
\end{equation}
Note that the summation here should exclude {self pairs and bonding pairs, corresponding to the terms} {when $i=j$ and $(|\alpha - \beta|-1)(\alpha - \beta)=0$ simultaneously hold.}
To describe the flexibility of the individual, we introduce the bending energy of the $i$-th individual as

\begin{equation}
        \label{equ:Bending}
	E^b_{i}=-\lambda\sum_{\alpha=1}^{n-2}{\bm p_{i,\alpha} \cdot \bm p_{i,\alpha+1}},
\end{equation}
where ${\bm p}_{i,\alpha} = ({\bm r}_{i,\alpha}-{\bm r}_{i,\alpha+1})/|{\bm r}_{i,\alpha}-{\bm r}_{i,\alpha+1}|$ denotes the orientation of the $\alpha$-th segment in the individual and $\lambda$ denotes the bending stiffness of the individual. The harmonic elastic energy accounting for the bond connecting two neighboring nodes in the $i$-th individual reads

\begin{equation}
        \label{equ:Spring}
	{E^s_{i}=\frac{k}{2}\left[ \sum_{\alpha=1}^{n-1}(r_{i,\alpha}^{i,\alpha+1}-r_0)^2 \right ]},
\end{equation}
where $k$ is the {spring} constant, and $r_0$ is the equilibrium length of the bond.

In our simulations, to describe the rotation of the cluster, we define an angular velocity
\begin{equation}
        \label{equ:w}
	\omega=\frac{1}{N}\frac{\sum_{i=1}^N \arcsin[||{\mbox{\boldmath $q$}_{i}( t) }\times {\mbox{\boldmath $q$}_{i}(t+\delta t)}||]}{\delta t},
\end{equation}
where
\begin{equation}
       \label{equ:q}
	\mbox{\boldmath $q$}_{i}=\frac{\mbox{\boldmath $r$}_{i}-\mbox{\boldmath $r$}_{\rm c}}{||\mbox{\boldmath $r$}_{i}-\mbox{\boldmath $r$}_{\rm c}||}
\end{equation}
is the unit orientational vector  pointing to the position of the $i$-th individual ${\bm r}_i =\sum_{\alpha=1}^{n}{\bm r}_{i,\alpha}/n$ from the geometric center of all the individual
\begin{equation}
       \label{equ:rc}
	\mbox{\boldmath $r$}_{\rm c}=\frac{1}{N}\sum_{i=1}^{N}\mbox{\boldmath $r$}_{i}.
\end{equation}
Meanwhile, we also introduce a chirality parameter
\begin{equation}\label{eq:ccc}
	c=\frac{||\sum_{i=1}^N (\bm r_i - \bm r_{\rm c}) \times \bm p_i||}{ N}
\end{equation}
to describe the spiral cluster structure. For a cluster with all the individuals precisely orienting towards the center, the chirality $c$ vanishes.

Besides, by using a typical active force and the number of particles in the cluster, we also define a typical angular velocity
\begin{equation}\label{eq:omega0}
	\omega_0=\frac{f_{\rm max}}{\xi R_{\rm c}N_{\rm node}},
\end{equation}
where $N_{\rm node}=nN$ is the total number of nodes (set as 1080 in our calculation).
\begin{table}[htbp]
        \centering
        \caption{Parameters used in simulations}
        \label{tab:arg}
        \resizebox{250pt}{!}{
        \begin{tabular}{|c|c|c|c|c|c|}
        \hline
		Variable & Meaning		   & Typical value& Range	&  Unit   \\
        \hline
		$\lambda$&  Bending rigidity       &    $1$   &$0,1,10$	& $\varepsilon$   \\
        \hline
		$D$      &  Diffusion coefficient &  $10^{-6}$  &$10^{-6},10^{-4},10^{-2}$ 	&  $\varepsilon/\xi$ \\
         \hline
		$n$      &  Number of nodes     	   &  $8$       &$8,10,12$  	& $ -$ \\
	\hline
		$N$      &  Number of individuals     	   &  $135$       &$135,108,90$  	& $ -$ \\
        \hline
		$f_{\rm max}$&  Maximum active force           &  $1.25$        &$1.25,2.5,5$ 	& $ \varepsilon/\sigma$\\
        \hline
		$R_{\rm c}$    &  Cutoff radius          &  $4$        &  - 	& $ \sigma$\\
        \hline
        $r_0$      &   Bond length      &$2^{1/6}$&    $0.1-1.0$ & $\sigma$\\
        \hline
        $ k$       &  Spring constant      &  $1$ &  $-$ & $ \varepsilon/\sigma^2$\\
        \hline
        $ \tau$    &  Time step            &  $1$ &  $-$ & $ \sigma^2\xi/\varepsilon$\\
        \hline
                \end{tabular}}
\end{table}
~
\begin{figure*}[htbp!]
        \includegraphics[width=500pt]{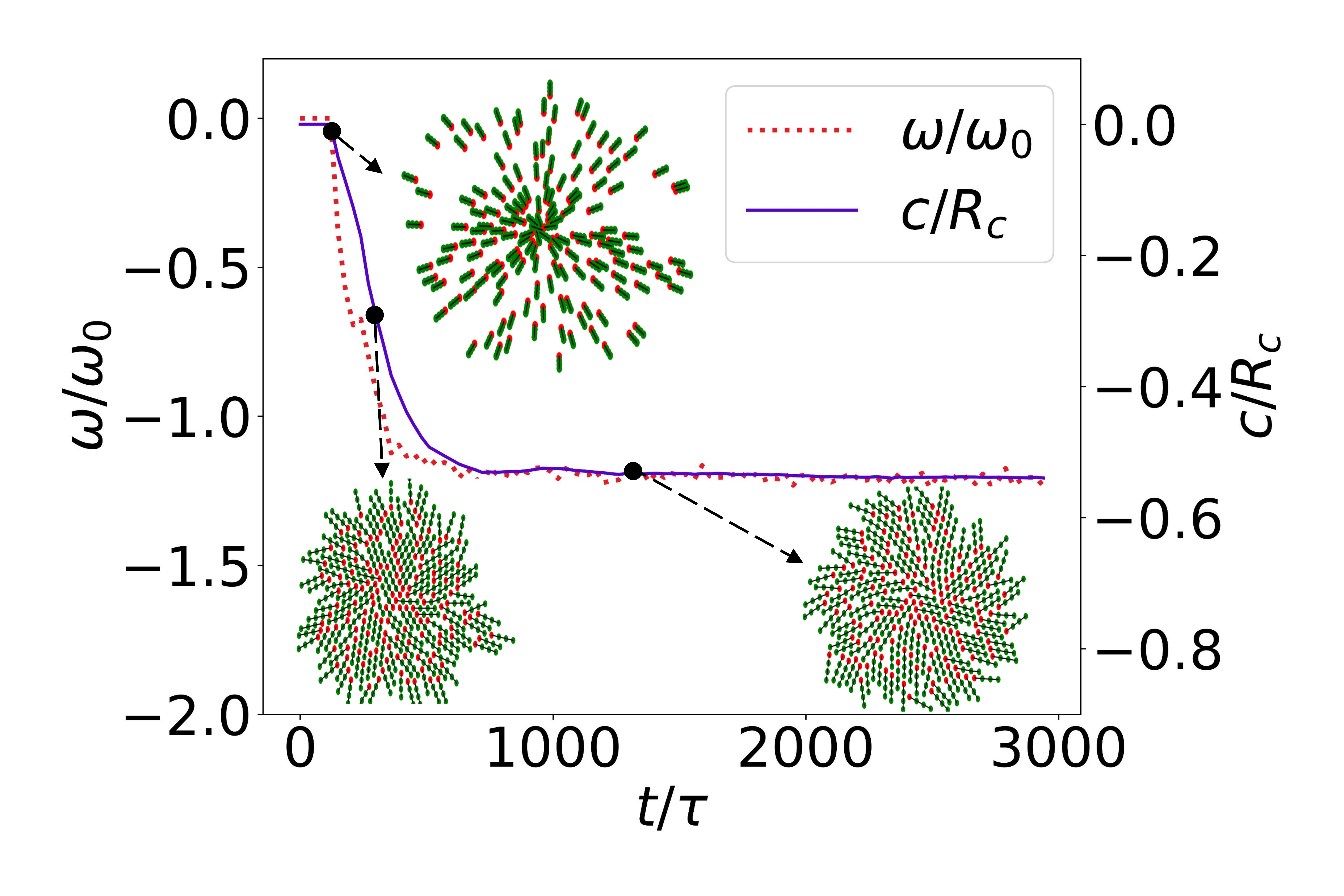}
        \caption{Temporal evolution of the cluster. The red curve shows the angular velocity of the cluster, and the purple curve shows the chirality. The negative sign of the angular velocity indicates a clockwise rotation. Insets show three snapshots of the cluster, with the head node of each individual colored in red. }
        \label{fig:timetr} 
\end{figure*}
\section{Results and Discussions}

We start our simulations with a cluster of individuals randomly positioned but not confined in a square box with parameters listed in Table I. {Scaled units are used for them so that the simulation results obtained can be universal. Typical values are used throughout the paper unless we want to investigate the dependence of the systems properties on the corresponding parameter.} Due to the possible food source in the square center known by the active individuals, they tend to flock toward the center{, a position of common interest.} All the individuals are assumed to have their initial orientations ${\bm p}_i$ pointing towards {the position of common interest} with a random perturbation $[-\pi/6,\pi/6]$ reflecting the difference of discrimination ability among the individuals, as shown in the top inset of Fig.~\ref{fig:timetr}. It is found that as time evolves, the initially dispersed individuals collide with each other and align their bodies either clock-wisely or counter-clock-wisely to form a rotating spiral cluster (insets in Fig.~\ref{fig:timetr}). The angular velocity $\omega$ and the chirality $c$ of the cluster quickly emerge and level off synchronizingly to a plateau (Fig.~\ref{fig:timetr}), indicating that a flocking-induced stable rotation rather than a transient one can be obtained.
In order to understand how the stable rotating cluster emerges from the collective motion of the active individuals, we performed simulations with different pairs of bending stiffness and bond length $(\lambda,r_0 )$. For each pair of the parameters, the simulations were repeated 80 times. For each simulation, if the angular velocity of the cluster (captured by the DBSCAN algorithm) converges to a certain value in a time of $3000\tau$ (as shown by the plateau in Fig.~\ref{fig:timetr}), a stable rotating cluster is confirmed and such a simulation is counted as one contribution of the probability to form a stable cluster (Fig.~\ref{phase}(a)). It is found that for a given $\lambda$, the probability experiences a sharp increase from 0 to 1 when the bond length $r_0$ increases to and crosses over a certain value (Fig.~\ref{phase}(a)). The $r_0-\lambda$ critical line is found to decrease and level off. In order to study the rotational behaviors of the cluster, using Eq.~(\ref{equ:w}), we calculate its angular velocity, which is obtained by taking its average over the last 1000 time points. In a similar manner as probability, there exists a critical line dividing the $r_0-\lambda$ space into two regions, the rotation region and the non-rotation region, as shown in Fig.~\ref{phase}(b). {As besides the elastic contribution and the steric interaction among the active individuals, each agent is assumed somehow to have the intelligence and thus the possibility to swerve its direction actively,} those with longer bond length are prone to change their force direction, and that enhances the possibility of swerving their movement towards the tangential direction. Such a tendency triggers the rotational motion of the individuals, which together with the steric interaction between the individuals, leads to a collective stable cluster rotation. This kind of rotation has to be closely related to the structural feature-chirality. The agreement of the critical line between Fig.~\ref{phase}(a) and Fig.~\ref{phase}(b) also supports Fig.~\ref{fig:timetr}. In a similar way, simulations were also performed for different bending stiffness and number of nodes $(\lambda, n)$. Once again a critical line dividing the rotation region and the non-rotation region is obtained, as shown in Figs.~\ref{phase}(c) and ~\ref{phase}(d).

\begin{figure}[h]
		\includegraphics[width=250pt]{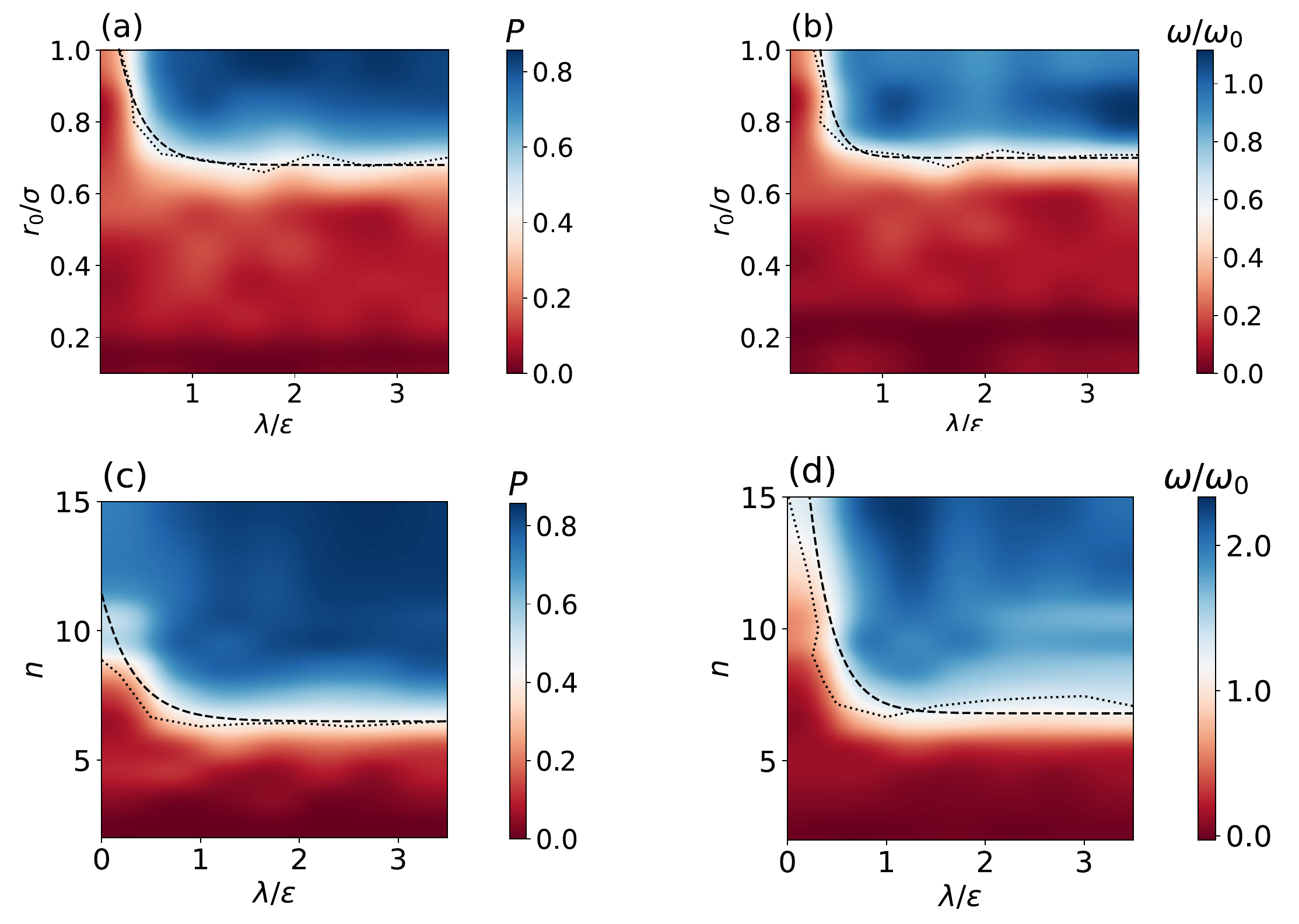}
	\caption{(a) The probability to form a stable rotating cluster in the $(r_0,\lambda)$ space. (b) The average angular velocity in the $(r_0,\lambda)$ space.(c) The probability to form a stable rotating cluster in the $(n,\lambda)$ space. (d) The average angular velocity in the $(n,\lambda)$ space. {The dotted lines are phase transition curves generated by using the contour function included in matplotlib, and the dashed lines are phase transition curves after exponential fitting.}}
         \label{phase}
\end{figure}

When a stable rotating cluster is formed, we notice a strong correlation between the angular velocity $\omega$ and the chirality $c$ for each simulation. Figure~\ref{fig:summary} shows the scattered plots of $(\omega/\omega_0,c/R_{\rm c})$ for the 128 simulations by changing one parameter while fixing the others as listed in Table~\ref{tab:arg}. When comparing the scattered plots for three different bending rigidities $\lambda$, it is found that the three straight lines for different bending rigidities almost collapse into a single straight line (Fig.~\ref{fig:summary}(a) left). The probability density functions (PDFs) for the angular velocity $\omega/\omega_0$ and the chirality $c/R_{\rm c}$ with a kernel-density-estimation (KDE) method demonstrate a bimodal distribution with two peaks located almost symmetrically with respect to $0$. The absolute peak positions for $\omega/\omega_0$ and $c/R_{\rm c}$ are both found to decrease with increasing $\lambda$ (Fig.~\ref{fig:summary}(a) middle and right ),
\begin{figure*}[h]
        \includegraphics[width=500pt]{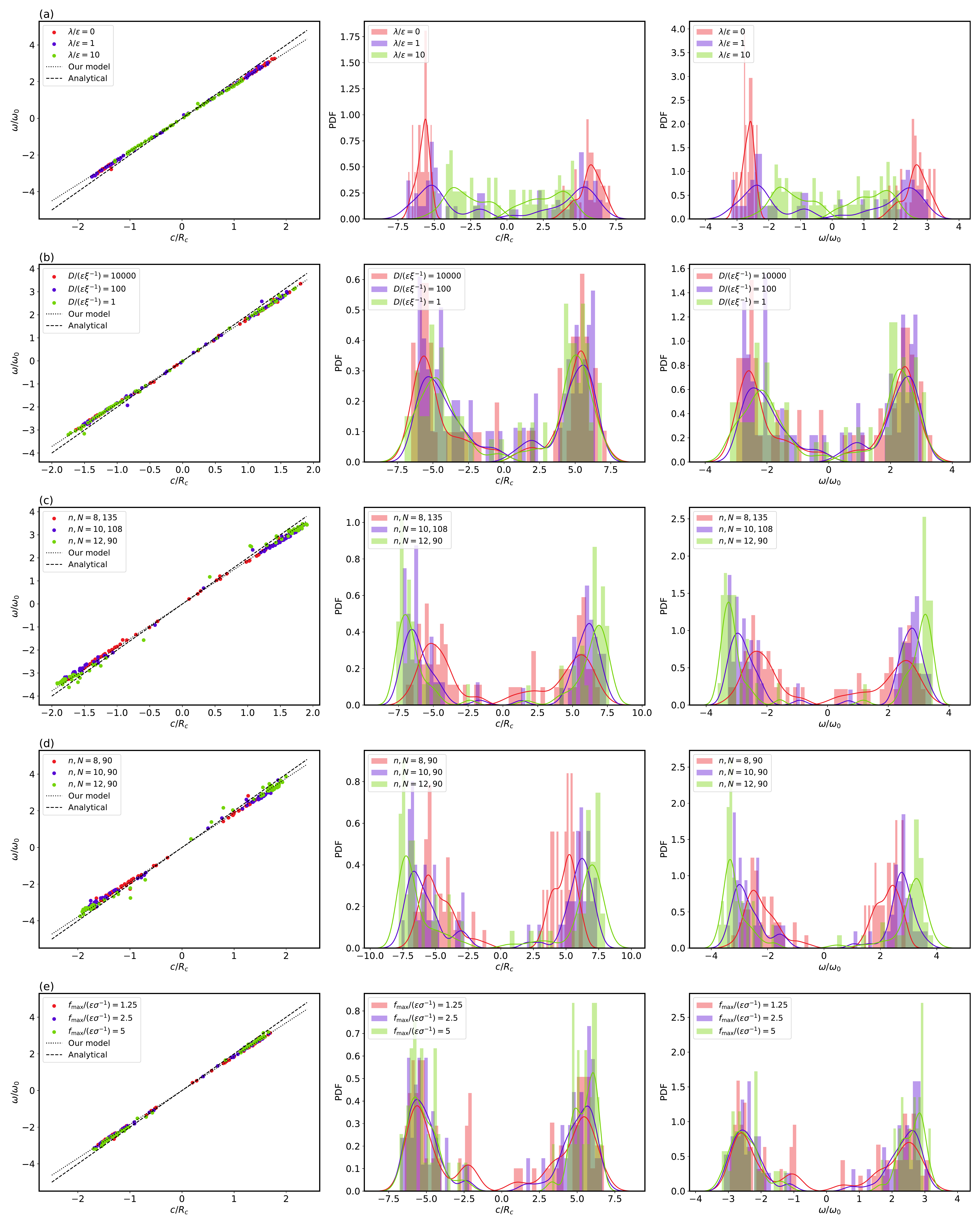}
	\caption{Scatter plots of $\omega/\omega_0-c/R_{\rm c}$ {(left column)}, and probability distribution functions {(The curves represent the distribution function obtained by the KDE method and the bar represents the histogram of the data)} ${\rm PDF}-c/R_{\rm c}$ and ${\rm PDF}-\omega/\omega_0$ at steady state as a result of repeated simulations for different (a) stiffness $\lambda$, (b) diffusion coefficient $D$, (c) number of node $n$, with $N_{\rm node}=1080$, (d) number of node $n$, with $N_{\rm node}=720,900,1080$ and (e) maximum active force $f_{\rm max}$. The dotted  lines and the dashed lines in $\omega/\omega_0-c/R_{\rm c}$ plots (left column) are based on our model {(Eq.~(\ref{eq:slope}))} and the theoretical analyses {(Eq.~(\ref{omegaanalytical}))}
, respectively.}
	\label{fig:summary}

\end{figure*}
indicating that the cluster formed by flexible individual rotates faster than that formed by rigid individual due to the chiral individual structure created by bending.

Besides the individual mechanical feature, we also explore the influence of diffusion on the angular velocity of the cluster. As shown in the left column of Fig.~\ref{fig:summary}(b), all the $(\omega/\omega_0,c/R_{\rm c})$ points at steady state, once again, collapse into a master straight line for different diffusion coefficients. {Even by increasing diffusion coefficient $D$ four orders of magnitude each time (Fig.~\ref{fig:summary} (b) middle and right), one sees just a slight increase of the absolute peak positions of the bimodal histogram distributions for normalized chirality $c/R_{\rm c}$ and angular velocity $\omega/\omega_0$, indicating that such a system is less subject to stochastic noise.}

 By varying the number of nodes $n$ for an individual and the total number of nodes $N_{\rm node}$ in {the system}, we find that they all result in a master straight line in the $(\omega/\omega_0,c/R_{\rm c})$ plane (Figs.~\ref{fig:summary}(c) and (d) left). Increasing the number of nodes or the total node number $N_{\rm node}$ enhances the chirality $c/R_{\rm c}$ (Figs.~\ref{fig:summary}(c) and (d) middle). The reason for the increase of $c/R_{\rm c}$ due to larger cluster size comes from the chirality definition Eq.~(\ref{eq:ccc}), which indicates that particles with larger distance from the cluster center contributes more to the chirality in the summation. The enhancement of chirality due to the increase of the number of nodes $n$ is supported by the phase diagram Fig.~\ref{phase} {if we consider the master linear relationship between angular velocity and chirality in Figs.~\ref{fig:summary}(c) and (d) (left).} On the other hand, increasing the number of nodes $n$ also leads to the increase of angular velocity $\omega$ (without changing $\omega_0$), while increasing the total number of nodes $N_{\rm node}$ decreases $\omega_0$ (Eq.~(\ref{eq:omega0})). As a result, they all lead to an increase of $\omega/\omega_0$ (Figs.~\ref{fig:summary}(c) and (d) right).

In a similar way, it is found that the maximum active force $f_{\rm max}$ does not change the slope of the normalized $(\omega/\omega_0,c/R_{\rm c})$ straight lines at the steady state. This is supported by the optimum values of $c/R_{\rm c}$ and $\omega/\omega_0$, which do not change with $f_{\rm max}$, as shown in the middle and the right figures of Fig.~\ref{fig:summary}(e).

By summarizing all the results above via non-dimensionalization, we obtain an empirical formula for the slope of the $\omega$-$c$ curve,
\begin{equation}
        \label{eq:slope}
	\frac{\omega/\omega_0}{c/R_{\rm c}}=1.86,
\end{equation}
 where $1.86$ is a dimensionless prefactor determined roughly by fitting. To rationalize this empirical formula, we propose a simplified model (see the Appendix) by approximating the density of individuals in the cluster as a uniform one and the cluster finally performs a rigid body rotation. The final angular velocity $\omega$ is determined by a torque balance between the active part and the dissipative part (see the Appendix), which is given by
 \begin{equation}
        \label{omegaanalytical}
	\frac{\omega/\omega_0}{c/R_{\rm c}}=2.
\end{equation}
{
 Assume that the cluster is composed of close packing chains during the flocking process, it can be divided into a core part with homogeneous node number distribution and an outer belt of inhomogeneous density of nodes, indicating that the node density can be given by
\begin{equation}
\label{eq:rg}
\rho(r) =
\begin{cases}
\dfrac{1}{d^2}  {~~~~\rm if~~} r<R \\
\dfrac{1}{d^2r}{\big [}R-\dfrac{R}{w}(r-R){\big]}  {~~~~\rm if~~} R<r<R+w,
\end{cases}
\end{equation}
where $d$ is the distance between neighboring nodes, $R$ is the radius of the core part, and $w$ is the width of the outer belt with inhomogeneous node density. Based on this picture, we can obtain (see the Appendix)
\begin{equation}
	\label{extended omega}
	\frac{\omega/\omega_0}{c/R_c}=\frac{2}{\zeta},
\end{equation}
where
\begin{equation}
\zeta(w/R)=1+2(w/R)+\frac{2}{3}(w/R)^2+\frac{1}{3}(w/R)^3,
\end{equation}
is a function of $w/R$, the ratio between the outer belt width and t
he radius of the core part. In the simulation $w/R$ is about 0.05, leading to the corrected prefactor of Eq.~(\ref{omegaanalytical}) as
\begin{equation}
2/\zeta(0.05)=1.815,
\end{equation}
which is almost equal to the simulation result.}
The spiral structure formation in our model is due to the alignment interaction between individual as a result of the steric interactions between the nodes. As two chain-like individual collide, they tend to align their bodies in the same direction. Given the initial condition we impose, the individual tend to tilt their directions in the same manner (clockwise or anti-clockwise) while they move towards the center, thus forming a spiral structure in which all the individual have their tail-to-head direction ${\bm p}_i$ deviated from pointing towards the center ${\bm r}_i - {\bm r}_{\rm c}$. In this way, the active force $f_i {\bm p}_i$ generates an active torque $({\bm r}_i - {\bm r}_{\rm c}) \times (f_i{\bm p}_i)$, which in turn rotates the cluster. A soft individual can generate a faster rotation because an individual can bend its body more, so that the tail-to-head direction ${\bm p}_i$ has a larger deviation from ${\bm r}_i - {\bm r}_{\rm c}$, thus generating a larger torque. We stress that not only the flexibility of the individual but also its bond length $r_0$ plays an important role in generating spontaneous stable rotation. To accomplish this, a suitable combination of individual length and flexibility is needed, as shown in the phase diagram Fig.~\ref{phase}. {The dependence of spontaneous stable rotation on chain length and flexibility might provide a new dimension in designing a robot cluster or a dynamic pattern of the drone swarm.}

\section{Conclusions}

In summary, we establish a model, which takes the intelligence of the individual into consideration, to account for the phenomenon of flocking-driven spontaneous stable rotation on a two-dimensional surface. A steady rigid-body-like rotation of the cluster is observed to synchronize with the formation of a spiral structure, which well explains the spontaneous stable rotation existing in nature formed by worms or sheep, when flocking together. The three important factors that influence the stable rotation are the active force, rigidity, and bond length related to the body orientation. The active force provides the driving element for the cluster, which turns out to produce an inherent angular momentum due to its chirality generated by flexibility and collectively enhanced by alignment interaction between individual. {This gives new insights into rotational ordering in active soft matter systems: bending rigidity produces chirality, and flocking towards an initially known position of common interest leads to spontaneous stable rotation.}

\section{ACKNOWLEDGMENTS}
We acknowledge financial support from National Natural Science Foundation of China under Grant No.12147142, No.11974292, No.12174323, and No.1200040838, Fundamental Research Funds for Central Universities of China under Grant No. 20720200072 (RM), and 111 project No.B16029.

\section{Appendix}
\subsection{Kernel density estimation}
We use the kernel density estimation (KDE) method to calculate the probability density function (PDF) of the angular velocity $\omega$ and the chirality $c$. The idea is to regard each sampling as a kernel function $K(x)$, and obtain the PDF through the superposition of the kernel functions
\begin{equation}
	\label{equ:kde}
	f(x)=\frac{1}{Nh}\sum_{i=1}^{N}K(\frac{x-x_i}{h}),
\end{equation}
where $1/{Nh}$ is the normalization factor, $x_i$ is the data of the $i$-th experiment, $N$ is the total number of experiments, $h=1.06 {\varrho} N_{\rm exp}^{-1/5}$ is the window width chosen by the rule of thumb, ${\varrho}$ is the standard deviation of all experimental data, and $N_{\rm exp}$ represents the number of repeated simulations. In order to make the estimated function smooth, the standard normal distribution function is used as the kernel function $K(x)$.
When we apply KDE to the entire data set of $\omega$ which typically demonstrates a bimodal distribution, the fitting PDF is much wider than the true distribution around each peak. We therefore apply KDE to the positive $\omega$ and negative $\omega$ separately and combine the two distributions to form the final PDF based on the ratio of the two data sets.

\subsection{A rigid-body rotation model}
In the simulations, we observe that as the cluster rotation finally reaches a steady state, the relative positions of the individual are almost unchanged and the rotation can be approximated as a rigid body rotation. The movement of each individual therefore obeys
\begin{equation}
	\dot{\tilde{\bm r}}_i =\omega {\bm z} \times {\tilde{\bm r}}_i,
	\label{eq:modle}
\end{equation}
where $\bm z$ is the unit vector pointing to the z-axis perpendicular to the motion plane and $\tilde{\bm r}_i$ represents the coordinate of the head node in the cluster. The force balance equation of each individual in the cluster reads
\begin{equation}
	n \xi \dot{\tilde{\bm r}}_i=\bm F_i^\alpha,
	\label{eq:balance}
\end{equation}
where $ n\xi \dot{\tilde{\bm r}}_i $ is the damping force of the $n$ nodes in a individual and $\bm F_i =$($f_{\rm max}/N_i$)${\bm p}_i$ is the active force generated by the head node. By substituting Eq.~(\ref{eq:modle}) into Eq.~(\ref{eq:balance}) and  taking the cross product of both sides with $\tilde {\bm r}_i$, we obtain
\begin{equation}
	\omega=\frac{||\sum_{i=1}^{N} \tilde{\bm{r}}_i \times (nf_{\rm max}/N_i)\bm{p}_i||}{\sum_{i=1}^{N} n\xi \tilde{ \bm{r}}_i^2} = \frac{M}{\Gamma}.
	\label{eq:w}
\end{equation}
Assuming the cluster has a constant individual node density $\rho_{\rm node}$, the number of nodes within the cutoff radius is $N_i = \rho_{\rm node} \pi R_{\rm c}^2$. The nominator $M$ in Eq. (\ref{eq:w}) represents the active torque, which can be reduced to
\begin{equation}
	M=\frac{cNnf_{\rm max}}{\pi R_{\rm c}^2 \rho_{\rm node}}.
	\label{eq:MM}
\end{equation}
The denominator $\Gamma$ in Eq.~(\ref{eq:w}) is the dissipative torque given by
\begin{equation}
\Gamma = \int_0^{R} 2\pi \xi \rho_{\rm node} r^3 dr = \frac{\pi\xi\rho_{\rm node} R^4}{2}.
\label{eq:Gamma}
\end{equation}
Using the equation $\rho_{\rm node} \pi R^2/n = N$ to replace $R$ in Eq.~(\ref{eq:Gamma}), and combining Eq.~(\ref{eq:MM}) and Eq.~(\ref{eq:Gamma}), we obtain Eq.~(\ref{omegaanalytical}).
If we keep the total number of nodes $N_{\rm node}=nN$ constant and define a typical frequency $\omega_0=f_{\rm max}/\xi R_{\rm c} N_{\rm node}$, the above equation leads to Eq.~(\ref{omegaanalytical}). {Given the inhomogeneous density distribution Eq.~(\ref{eq:rg}), the dissipative torque is written as
\begin{equation}
\label{newGamma}
\Gamma =\sum_i \xi r_i^2 = \int_0^R 2\pi \xi \frac{1}{d^2} r^3 {\rm d}r +\int_R^{R+nd} 2\pi \xi \frac{1}{d^2r}{\big[}R-\frac{R}{w}(r-R){\big]} r^3 {\rm d}r.
\end{equation}
Using Eqs.~(\ref{eq:rg}) and (\ref{newGamma}), Eq.~(\ref{omegaanalytical}) becomes Eq.~(\ref{extended omega}).
}

\end{document}